\documentclass[usenatbib]{emulateapj}
\slugcomment{Submitted to the Astrophysical Journal, 2015 May 28}  
\pdfoutput=1
\shorttitle{Strong Lenses in Forthcoming Optical Surveys}
\shortauthors{Collett (2015)}

\usepackage{color}			      
\definecolor{midgray}{gray}{0.4}		
\definecolor{orange}{rgb}{1,0.5,0}    

\usepackage{xspace}
\usepackage{graphicx}
\usepackage{amsmath}

\def\spose#1{\hbox  to 0pt{#1\hss}}  
\newcommand{\lta}{\mathrel{\spose{\lower 3pt\hbox{$\sim$}}\raise  2.0pt\hbox{$<$}}}
\newcommand{\gta}{\mathrel{\spose{\lower  3pt\hbox{$\sim$}}\raise 2.0pt\hbox{$>$}}}

\newcommand{\be}{\begin{equation}}
\newcommand{\ee}{\end{equation}}

\newcommand{\citepeg}[1]{\citep[e.g.][]{#1}}

\newcommand{\kms}{\ifmmode  \,\mathrm {km}\,s^{-1} \else $\,\mathrm{ km\,s^{-1} } $ \fi }
\newcommand{\kpc}{\ifmmode  {\mathrm{kpc}}  \else ${\mathrm{  kpc}}$ \fi  }  
\newcommand{\pc}{\ifmmode  {\mathrm{ pc}}  \else ${\mathrm{ pc}}$ \fi  }  
\newcommand{\Msun}{\ifmmode {\mathrm{ M_{\odot}}} \else ${\mathrm{ M_{\odot}}}$ \fi} 
\newcommand{\Zsun}{\ifmmode {\mathrm{{Z_{\odot}}}} \else ${\mathrm{ Z_{\odot}}}$ \fi} 
\newcommand{\yr}{\ifmmode yr^{-1} \else $yr^{-1}$ \fi} 
\newcommand{\hMsun}{\ifmmode h^{-1}\,\rm M_{\odot} \else $h^{-1}\,mathrm{\ M_{\odot}}$ \fi}

\def\spose#1{\hbox  to 0pt{#1\hss}}  
\renewcommand{\lta}{\mathrel{\spose{\lower 3pt\hbox{$\sim$}}\raise  2.0pt\hbox{$<$}}}
\renewcommand{\gta}{\mathrel{\spose{\lower  3pt\hbox{$\sim$}}\raise 2.0pt\hbox{$>$}}}

\renewcommand{\be}{\begin{equation}}
\renewcommand{\ee}{\end{equation}}

\newcommand{\bea}{\begin{eqnarray}}
\newcommand{\eea}{\end{eqnarray}}

\def\ringfinder{{\sc Ringfinder} }

\newcommand{\comment}[1]{}
\newcommand{\comments}[1]{}

\def\icg{Institute of Cosmology and Gravitation, University of Portsmouth, Burnaby Rd, Portsmouth, PO1 3FX, UK}

\def\collettemail{\tt thomas.collett@port.ac.uk}

\begin{document}

\title{The population of galaxy-galaxy strong lenses in forthcoming optical imaging surveys}

\author{Thomas E. Collett}
\affil{\icg}
\email{\collettemail}





\begin{abstract}
Ongoing and future imaging surveys represent significant improvements in depth, area and seeing compared to current data-sets. These improvements offer the opportunity to discover up to three orders of magnitude more galaxy-galaxy strong lenses than are currently known. In this work we forecast the number of lenses discoverable in forthcoming surveys and simulate their properties. We generate a population of statistically realistic strong lenses and simulate observations of this population for the Dark Energy Survey (DES), Large Synoptic Survey Telescope (LSST) and Euclid surveys. We verify our model against the galaxy-scale lens search of the Canada-France-Hawaii Telescope Legacy Survey (CFHTLS), predicting 250 discoverable lenses compared to 220 found by \citet{ringfinder}. The predicted Einstein radius distribution is also remarkably similar to that found by  \citet{sonnenfeld3}. For future surveys we find that, assuming Poisson limited lens galaxy subtraction, searches in DES, LSST and Euclid datasets should discover 2400, 120000, and 170000 galaxy-galaxy strong lenses respectively. Finders using blue minus red ($g-i$) difference imaging for lens subtraction can discover 1300 and 62000 lenses in DES and LSST. The uncertainties on the model are dominated by the high redshift source population which typically gives fractional errors on the discoverable lens number at the tens of percent level. We find that doubling the signal-to-noise ratio required for a lens to be detectable, approximately halves the number of detectable lenses in each survey, indicating the importance of understanding the selection function and sensitivity of future lens finders in interpreting strong lens statistics. We make our population forecasting and simulated observation codes publicly available so that the selection function of strong lens finders can easily be calibrated.

\end{abstract}

\keywords{gravitational lensing: strong}

\setcounter{footnote}{1}


\section{Introduction}
\label{sec:intro}

Strong gravitational lensing by galaxies can be used to probe both astrophysics and cosmology. To date several hundred galaxy-galaxy strong lenses have been discovered in heterogeneous searches of photometric and spectroscopic survey data \citep{class, sl2s, slacs, inada, negrello}. The sample has been used to constrain the masses and density profiles of galaxies \citep{auger2010,barnabe}, the dark sub-halo population \citep{vegetti2014}, cosmological parameters \citep{collett2014,oguri2012,suyu2014}, the high redshift luminosity function \citep{bn2015} and the nature of high redshift sources \citepeg{slacsXI, quider}. 
For many of these analyses the shortage of suitable strong lenses is a major limiting factor. 

Several ongoing, and near-future, wide and deep sky surveys offer improved depth, area and resolution compared to existing data \citep{DES,HSC,LSST,Euclid}. These surveys have the potential to increase the current galaxy-scale lens sample by orders of magnitude \citep{kuhlen2004, marshall2005}. With a large increase to the known strong lens population, current work could be extended to new regimes; lower lens masses, higher redshift lenses and intrinsically fainter sources. In turn this will allow for investigations into the luminosity and redshift trends of lens and source properties. For example, investigating trends in the mass-to-light ratio of galaxies, the distribution of dark-matter in galaxies and the dark-substructure population will much more tightly constrain the nature of dark matter, the initial mass function and galaxy formation physics than is currently possible \citepeg{vegetti2014,sonnenfeld3}.

Future surveys also have the potential to discover a population of exotic --and rare-- strong lenses systems. For example, compound lenses are powerful probes of dark matter \citep{sonnenfeld2012} and cosmological parameters \citep{collett2012} as are strongly-lensed supernovae \citep{refsdal} and strong lensing catastrophes \citep{orban}, but only a handful of these systems are known \citep{gavazzi,SNrefsdal,SNquimby}, which limits the precision of cosmological constraints achievable with strong lensing \citep{collett2014}.

Additional to science using sub-samples of lenses, the size and properties of the full ensemble of strong lenses is dependent on both cosmological parameters and the source and deflector populations \citepeg{oguri2012}; future surveys will provide a much more powerful sample for strong lensing statistics.

Recent analytical work has forecast the number of lensed point sources likely to be discovered in spectroscopic surveys \citep{serjeant2014} and in photometric surveys \citep{om10}, however making forecasts for lenses with extended background sources requires one to move beyond analytic analyses due to differential magnification across the source plane. \citet{dobler2008} simulated lensing of spectral line emission regions and then simulated observing such objects with the SDSS spectrograph to model the selection function in the SLACS survey. {By changing the source population to galaxies \citep{ilbert} and the transfer function to that of forthcoming surveys \citepeg{chang2015} the methodology of \citet{dobler2008} can be extended to predict galaxy-galaxy strong lensing rates in forthcoming photometric surveys.}

The goal of this work is to answer the question {\it``How many galaxy-galaxy lenses can forthcoming imaging surveys potentially discover?''} To answer this question we first build a population of realistic strong lenses (Section \ref{sec:population}) and then simulate observations of these lenses (Section \ref{sec:observations}) for four recent and forthcoming photometric surveys; CFHTLS, DES, LSST and Euclid. Of course actually discovering strong lenses in these surveys will require the development of new methods and algorithms \citepeg{ringfinder,brault,kueng}, which is beyond the scope of this article, but this work is intended to motivate such developments by providing both realistic expectations for lens finders and realistic simulations to test lens finders upon. We also investigate how the features of lens finders can affect the number of strong lenses that they can discover, paying attention to how they alter the effective seeing and signal-to-noise of the data. In Section \ref{sec:errors} we discuss our results in the context of the uncertainties inherent to how we have constructed the lens population and how we have assumed future lens finders might work. We summarize and conclude our results in Section \ref{sec:conclude}. Throughout this paper we assume a flat $\Lambda$CDM cosmology with $\Omega_M=0.7$, and $h=0.7$.


\section{Creating a population of realistic mock galaxy-galaxy strong lenses}
\label{sec:population}
Strong  gravitational lensing occurs when a massive foreground object is sufficiently well aligned with a background source that the lens equation has multiple solutions and hence multiple images of the source can form. The population of strong lenses therefore depends on the population of massive objects, background sources, and the geometry of the Universe. Once the lens population is known individual strong lens systems can be drawn from the population and simulated to assess if strong lensing detectable. Simulating individual systems requires knowledge of the light profile of the lens and source, the density profile of the source and the angular diameter distances between observer, lens and source.

\subsection{The foreground deflector population}
\label{deflectors}

Observations have shown that the mass profiles of elliptical galaxies are well approximated by isothermal mass distributions \citepeg{auger2010}. In this work we will consider only strong lensing by elliptical galaxies, which dominate the galaxy-galaxy lensing cross section \citep{om10}, and then assume that all of the foreground deflectors are singular isothermal ellipsoids (SIEs). The density profile of an SIE is given by
\begin{equation}
\rho(\tilde{r})~=~{{\sigma_V}^2 \over {2 \pi G \tilde{r}^2}}\label{SIS}.
\end{equation}
where $\sigma_V$ is the velocity dispersion of the lens, $\tilde{r}$ is the elliptical distance from the center ($\tilde{r}^2 = x^2/q + qy^2 $), $q$ is the ratio of semi-major and semi-minor axes( i.e. the flattening).
Einstein radii of SIEs are given by the stellar velocity dispersion of the galaxy, $\sigma_V^2$, and angular diameter distances between observer, lens and source ($D_{ij}$)
\be
\theta_{\text{E}}^{\text{SIS}}~=~{4 \pi{{\sigma_V^2}\over{c^2}}{{D_{\text{ls}}} \over {D_{\text{s}}}}}.
\ee
We neglect the lensing contribution of matter along the line-of-sight.

For the light profile of the deflector, we assume  an elliptical de Vaucolours profile concentric with, and aligned with, the mass \citep[e.g. as observed by][]{gavazzi,zuzzana}. We make the approximation that the effective radius, $R_e$, of the deflector is the same in all observed bands, but allow the absolute magnitude $M$ in each band to vary. For the colors of the lens we assume the rest-frame SED of a passive galaxy whose star formation was a single burst 10 Gyrs ago.

Given these simplifying assumptions, our model for the deflector is thus described by five parameters; the lens redshift, $\sigma_V$, $q$, $R_e$, and $M_r$. Rotational, and translational symmetry allow us to place the deflector at the center of the coordinate system and align the semi-major and coordinate x-axes. 

\citet{sdssvdisp} used Sloan Digital Sky Survey \citep[SDSS;][]{sdss} data to derive the velocity dispersion function of elliptical galaxies in the local Universe, finding
\begin{equation}
{\text{d}n}~=~ \phi_* \left({\sigma \over \sigma_*}\right)^\alpha \exp \left[-\left({{\sigma}\over{\sigma_*}}\right)^\beta \right] {\beta \over \Gamma(\alpha/\beta)}{\text{d}\sigma \over \sigma},
\label{dndsig}
\end{equation}
where $\phi_*~=~8.0 \times 10^{-3}h^3 $Mpc$^{-3}$, $\sigma_*~=~161$ kms$^{-1}$, $\alpha~=~2.32$ and $\beta~=~2.67$. In our model we assume that neither the shape nor the normalization of this function vary with redshift; this is consistent with the observations of \citet{bezanson}.

From Equation \ref{dndsig} we can draw a velocity dispersion for each deflector to be simulated and a redshift from the differential co-moving volume function ${\text{d} \text{Vol}/\text{d} z}$. The other parameters of our lens model,  $q$, $R_e$ and $M$ are covariant with the redshift and velocity dispersion. The fundamental plane of elliptical galaxies allows us to infer $R_e$ and $M$ given $\sigma_V$; we use the functional form and scatter as derived in \citet{HydeandBernardi}.
The final parameter is the flattening, which is known to correlate with $\sigma_V$; more massive galaxies tend to be closer to spherical than less massive ones. This trend is clearly seen in the SDSS data, which we fit with a Rayleigh distribution, where the Rayleigh scale parameter $s$ is linear in $\sigma_V$
\be
P\left(1-q| s=(A + B \sigma_V)\right)=\frac{1-q}{s^2}\exp{\left(\frac{-(1-q)^2}{2 s^2}\right)}
\ee
The fit values are $A = 0.38$ and $B = 5.7 \times 10^{-4} \left(\mathrm{km s^{-1}})^{-1}\right)$. To exclude highly-flattened mass profiles, we truncate the Rayleigh distribution at $q=0.2$.

\begin{figure*}
  \centering
    \includegraphics[width=0.32\textwidth,clip=True]{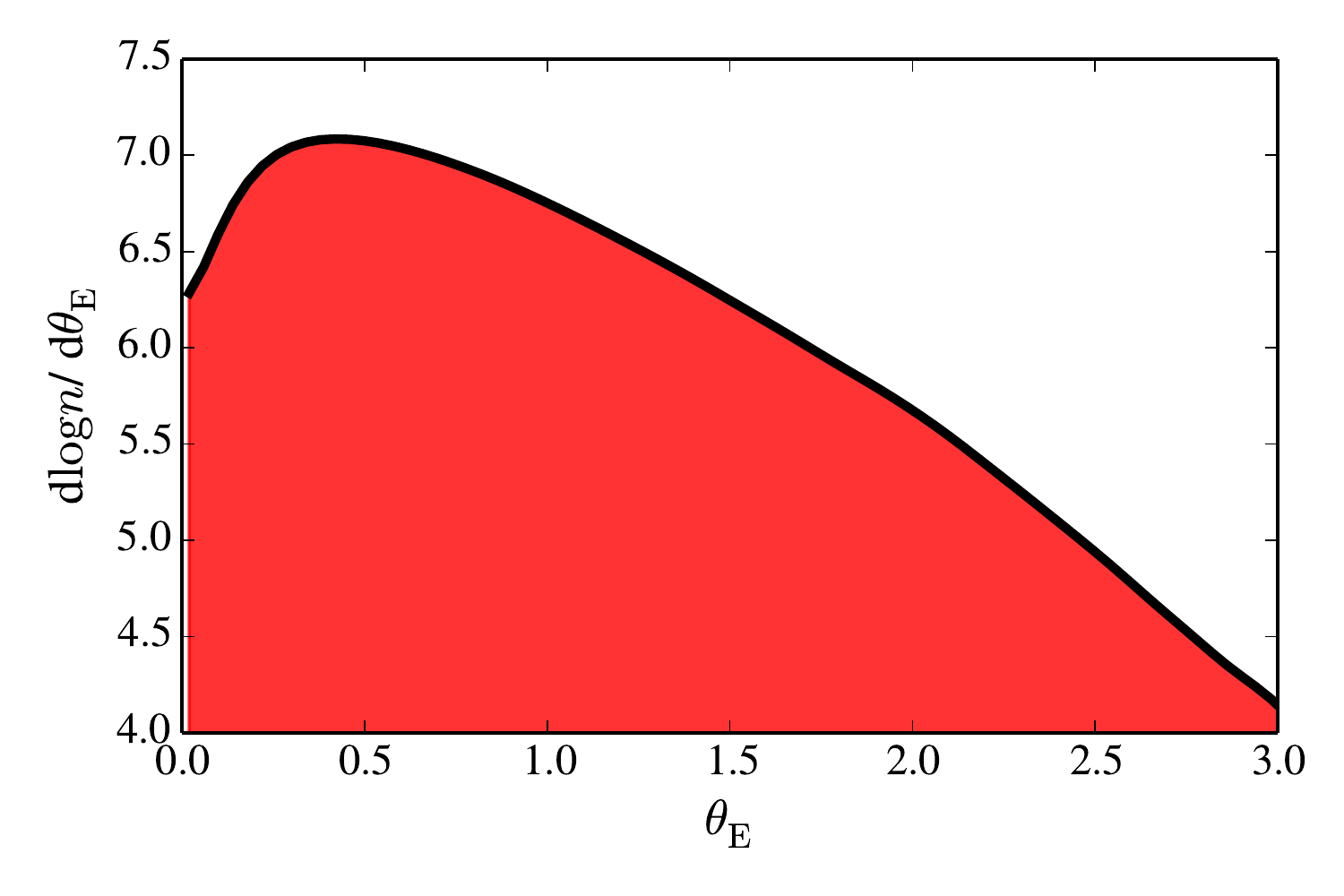}
    \includegraphics[width=0.32\textwidth,clip=True]{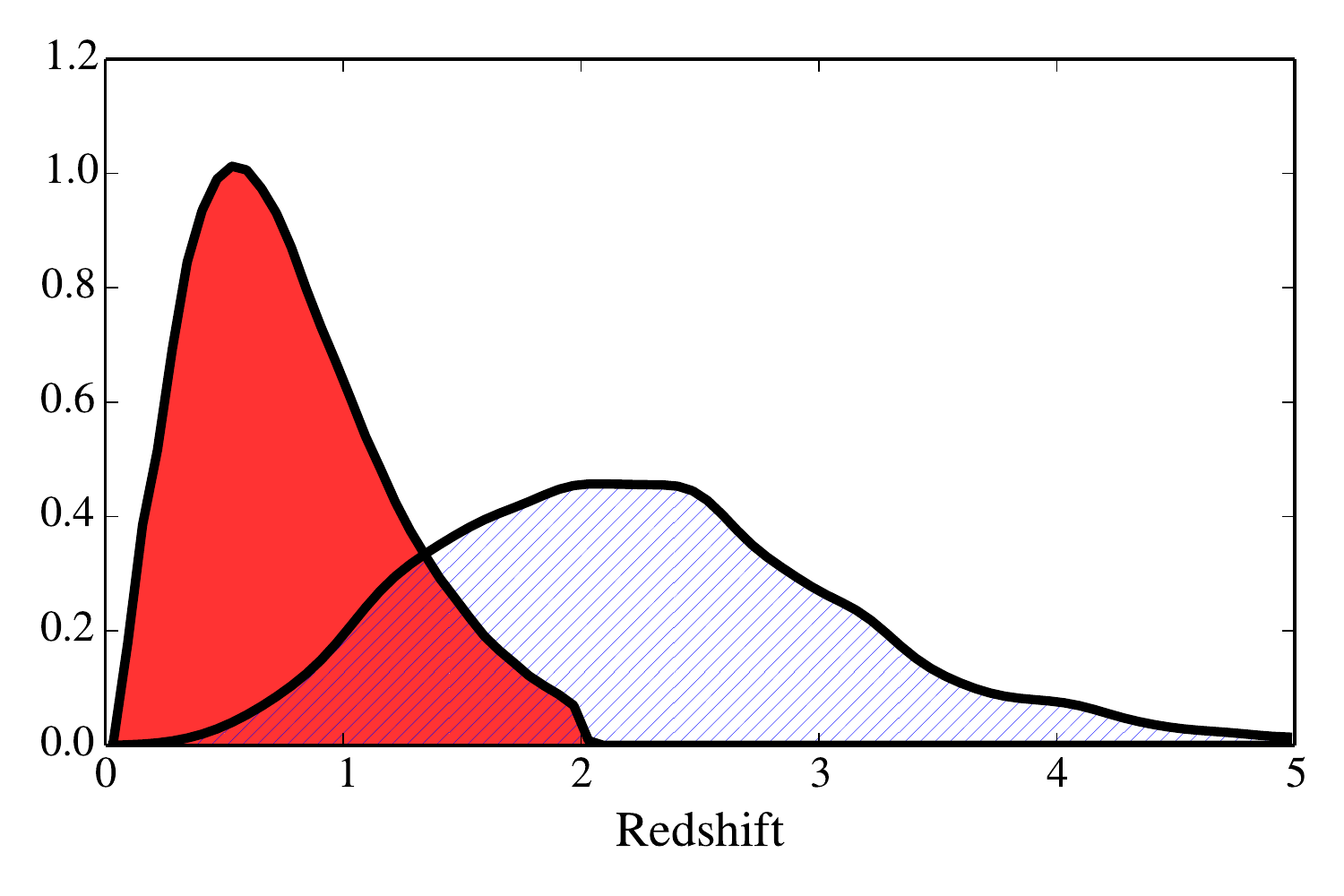}
    \includegraphics[width=0.32\textwidth,clip=True]{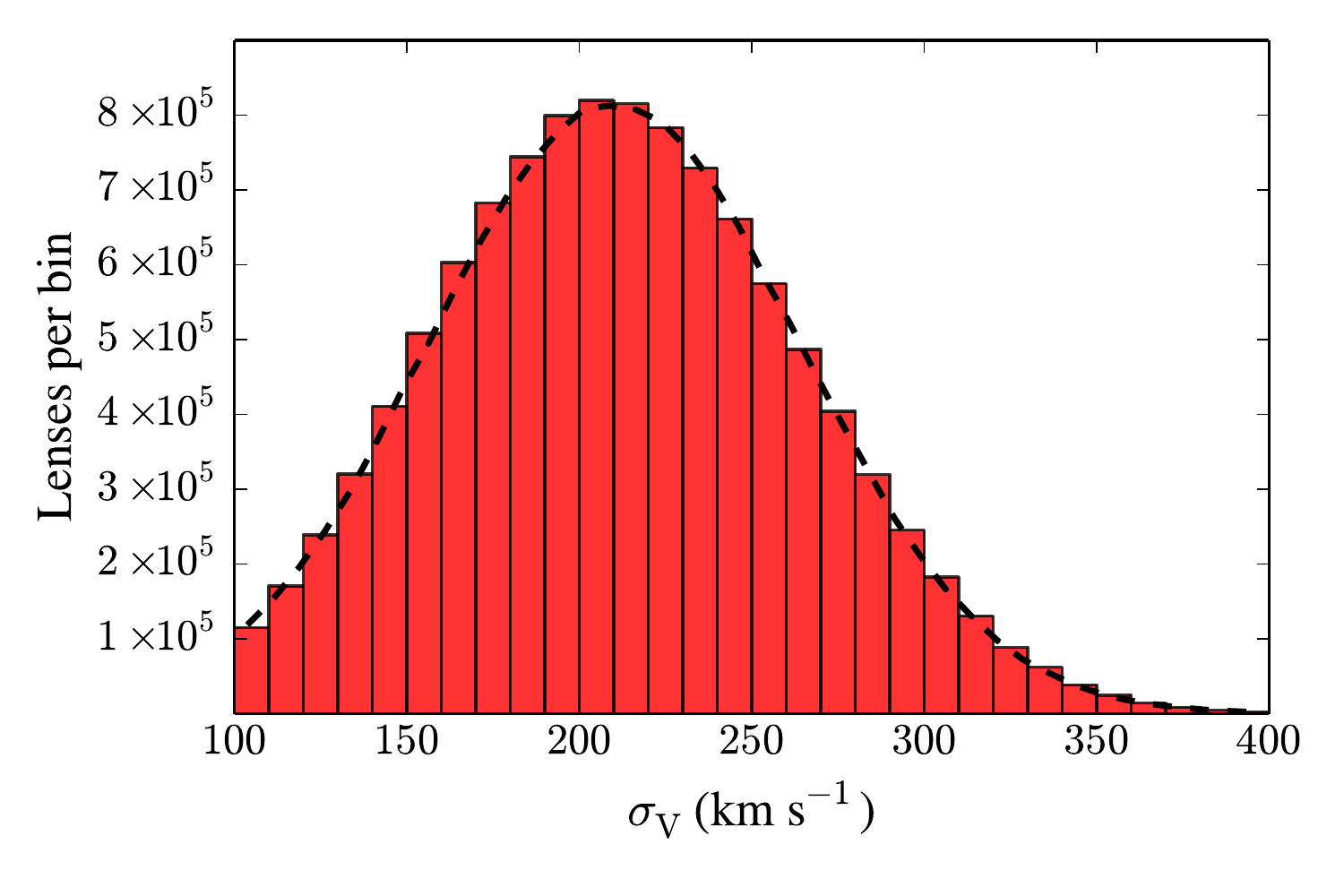}
\caption {The properties of the ideal lens sample i.e. all galaxy-galaxy lenses in the Universe with a deflector with $\sigma_V>100$km/s and a source with unlensed $i<27$. The left panel shows d$\log n$/d$\theta_\mathrm{E}$. The middle panel shows the distribution of lens redshifts as the solid line and the distribution of source redshifts as the dashed line. The right panel shows the number of lenses in each $\sigma_V$ bin of width 10 km/s and the dashed line shows a Gaussian fit to this distribution, with mean 210 and width 54 km/s.}
\label{fig:idealproperties}
\end{figure*}

\subsection{The background source population}
\label{sources}

For the lensed background sources we need only know their redshifts, light profiles, colors and absolute magnitudes. Since typical lensed sources are at high redshift, the light profiles and spectroscopic redshifts of the unlensed source population is often beyond the capability of current telescopes. Neglecting intrinsically faint, highly magnified sources could potentially cause us to significantly underestimate the detectable strong-lens population; we therefore use the sky catalogs simulated for the LSST collaboration by \citet{connolly}. The \citet{connolly} catalog is generated by ray-tracing through the Millennium dark matter simulation \citep{springel2005}, with galaxies added using the semi-analytic prescription of \citet{delucia} adjusted to mimic the luminosity and color distributions of low redshift galaxies and the redshift and number-magnitude distributions derived from deep imaging and spectroscopic surveys.  The final catalog comprises a 4.5x4.5 square degree footprint on the sky and samples halo masses down to $2.5 \times 10^{9}$\Msun. Due to the resolution limit of the simulations, this catalog is only complete down to $i \sim 27.5$, which may be insufficiently deep to perfectly reconstruct the faintest lens population of LSST but sufficient for the other surveys that we investigate here.

For all sources, we assume that the light profile is given by an elliptical exponential profile with ellipticity drawn from a Rayleigh distribution with scale parameter $s=0.3$, truncated at $q=0.2$. For the effective radius of the profile we assume
\be
\label{eq:size}
\log_{10}\left(\frac{R_e}{\mathrm{kpc}}\right)=\left(\frac{M_V}{-19.5}\right)^{-0.22} \left(\frac{1.+z}{5.}\right)^{-1.2}+N(0.3)
\ee
where $N(0.3)$ is a Gaussianly distributed random variable of scatter 0.3. The relation and scatter are in line with the results of \citet{mosleh} and \citet{huang} which both show trends towards more compact sources at higher redshifts.

\subsection{The properties of the realistic lens systems, before observing with a telescope}

Given the lens and source populations generated in sections \ref{deflectors} and \ref{sources}, we can project the lensing cross section back onto the source population, giving the expected distribution of lens and source properties for an idealized survey. Our ideal model predicts 11 million lenses on the whole sky, with a lens velocity dispersion $\sigma_V>100$km/s and a background source with $i<27$. The Einstein radius distribution, deflector velocity dispersion distribution and the source and lens redshift distributions are shown in Figure \ref{fig:idealproperties}. The model predicts that the population of strong lenses is dominated by galaxies with $\sigma_V \approx 200$ km/s and the distribution is well approximated by a Gaussian with mean 210 km/s and one sigma width of 54 km/s. The majority of lenses have Einstein radii around 0.5 arcseconds and sources close to the completeness limit of the source catalog. Even after strong lensing magnification most sources are too faint to be detectable by Euclid, and the Einstein radii are too small to be resolvable with LSST.


\section{Simulated observations of strong lenses with survey telescopes}
\label{sec:observations}

\begin{figure*}
    \includegraphics[width=\textwidth,clip=True]{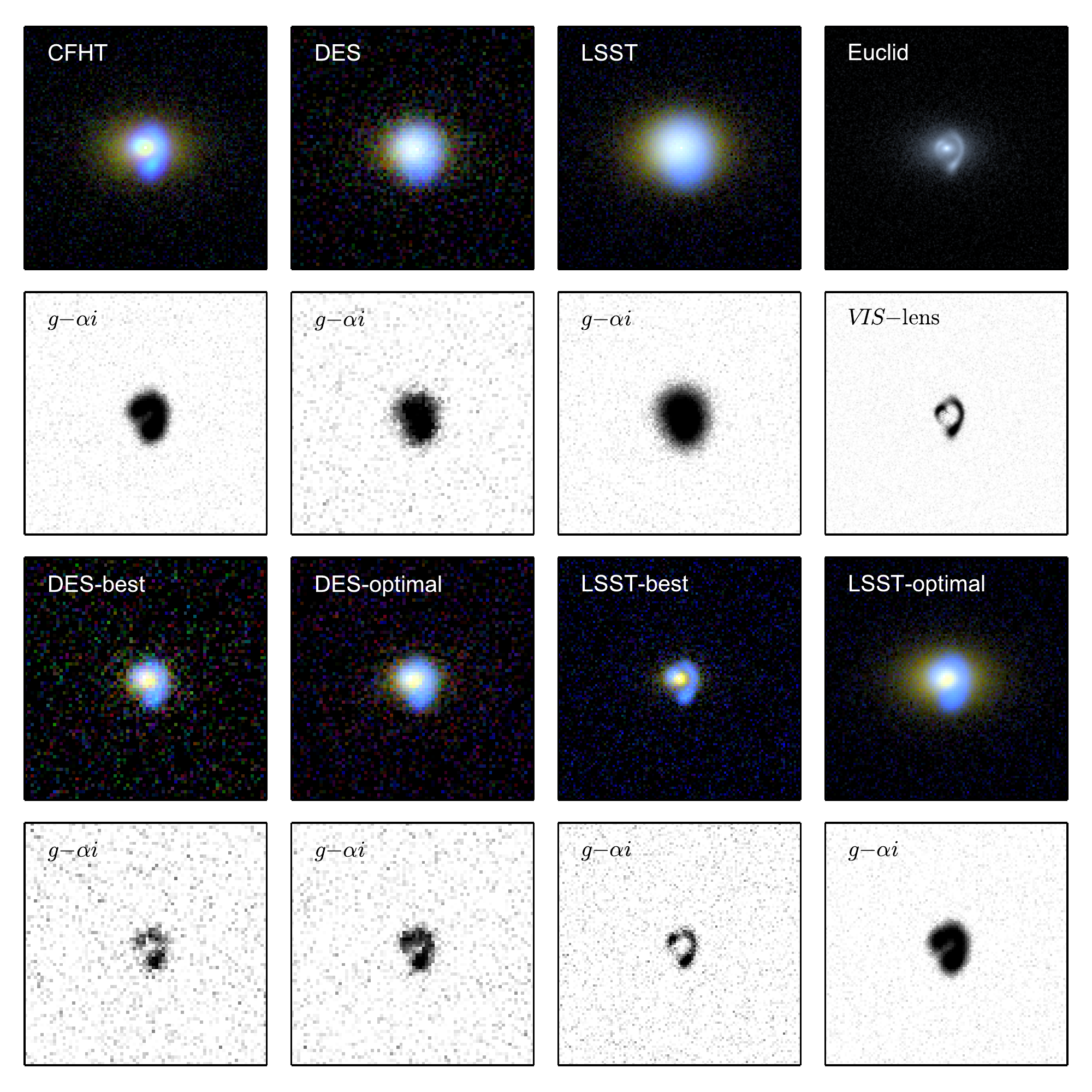}
    \caption {A simulated lens as would be observed by forthcoming telescopes. From left to right, the top row shows  a $gri$ color composite image for the full-depth co-add of CFHT, DES and LSST. The top right shows the Euclid image. From left to right the second row shows the $g-\alpha i$ difference image corresponding to the color composite above. Since Euclid has only one optical band, the second row far right shows the VIS residual assuming the galaxy can be subtracted perfectly (although the noise on the galaxy light remains). From Left to right, the third row shows the $gri$ composites for the best single epoch DES imaging, the optimally stacked DES imaging, the best single epoch LSST imaging, and the optimally stacked LSST imaging. The fourth row shows the $g-\alpha i$ difference image corresponding to the color composite above. Our signal-to-noise detectability criteria are satisfied in all images, but we do not consider the lens to be detectable in the CFHT, DES and LSST full co-adds as the arcs are not resolved. The CFHT full-stack only marginally fails the resolution criteria, whilst the optimally stacked LSST image only marginally satisfies it}
    \label{fig:examplelens}
\end{figure*}

In practice it is impossible to observe the idealized noiseless, unconvolved lenses produced by our model; real observations of strong lenses will involve the convolution with a point spread function, and various sources of noise. To investigate the discoverable lens population in future surveys, we must therefore include the transfer function of each survey telescope. We simulate observations by CFHTLS, DES, LSST and Euclid.

Our prescription for the transfer function is a simple one, but it encapsulates all the significant features relevant to lens finding. For each survey, the ideal lenses are evaluated on a pixellated grid with pixel-scale equal to the detector pixel-scale\footnote{{We subsample the pixel grid by a factor of 25 in regions where the source profile varies quickly}}, then convolved with the circularly symmetric PSF discretized to the sample pixel-scale. We then simulate the noise assuming Poisson noise from the lens, source and a uniform sky background and a constant read-noise for each exposure. We neglect cosmic-rays, artifacts and ghosts. For complete and on-going surveys, the seeing, sky-brightness, zero-points, exposure-times, number of exposures, pixel-scale, read-noise, filter set and survey area are taken from the existing observations. For future surveys we use the LSST observation simulator \citep{connolly}, and the Euclid design parameters. The assumed survey parameters are summarized in Table \ref{table:surveys}.

The seeing and sky-brightness are stochastic parameters that vary significantly over the course of a survey, hence  for each simulated exposure we draw from the real (or expected) distribution of seeing and sky-brightness observed {\it in each filter set}. This point is significant to the expected number of discoverable strong lenses for two reasons. Firstly, survey strategies are optimized to take bluer-band data in moonless skies and redder-band data when the moon is up. Secondly, survey strategies are often optimized for specific science goals, for example DES preferentially takes $r$- and $i$-band data in good seeing to improve the weak lensing analysis \citep{DES}. Survey strategies  therefore lead to seeing and sky-brightness distributions that are often strongly correlated, and not random draws from the seeing and sky brightness distributions of the telescope site. The most significant feature for lens finding is the strategy used for allocating good and bad seeing time. Since the majority of background sources are blue \citep{ringfinder} and the Einstein radius distribution of our lenses peaks at $\sim$0.5 arcseconds, strategies that penalize $g$-band seeing are likely to limit the potential for discovering strong lenses.

In Figure \ref{fig:examplelens} we show an example of a simulated lens as would be observed with each of CFHT, DES, LSST and Euclid.

\begin{table*}
 \caption{Properties of large area imaging surveys considered in this work. 
These are fiducial  numbers based on assumptions we describe in the text; given are survey
  area $\Omega$, effective collecting area, filter-set,  median seeing in each filter (FWHM), median sky brightness in each filter, total exposure time per filter, and pixel-scale.
\label{table:surveys}} 
 \begin{tabular}{@{}lccccccccc}
  \hline
Survey & $\Omega$  & Collecting area$^a$ &  Filters & Seeing$^b$ & Sky$^b$ & Exposure time &Pixel-scale \\                
       & [deg$^2$] & [m$^2$] &  & [arcsec] &  & [s] & [arcsec] \\                
 \hline
CFHTLS& 150   & 8.8 & \{$g,r,i^c$\} &\{0.83, 0.79, 0.69\}&\{21.7, 20.8, 19.7\}&\{3500, 5500, 5500\}  & 0.187 \\
DES   & 5000  & 9.6 & \{$g,r,i^c$\} &\{1.24, 1.05, 0.96\}&\{21.7, 20.6, 19.4\}&\{900, 900, 900\}     & 0.263 \\
LSST  & 20000 & 35  & \{$g,r,i^c$\} &\{0.81, 0.77, 0.75\}&\{21.7, 21.1, 20.0\}&\{3000, 6000, 6000\}  & 0.18  \\
Euclid& 15000 & 1.0 & \{$VIS^d$\}   &\{0.18\}            &\{22.2$^e$\}            &\{1610$^e$\}              & 0.1   \\
 \hline
 \end{tabular}
\newline\footnotesize
\flushleft{$^a$ We do not directly use the collecting area in our analysis, we instead use the instrumental zero-points for each survey for CFHT \citep{megacam} and DES \citep{decam}, for LSST we rescale the DES zero-points by the ratio of collecting areas. For Euclid we assume a zero-point of 25.5 AB (S. Niemi, Private communication)}\vspace{-0.5\baselineskip}
\flushleft{$^b$ Median quantities shown}\vspace{-0.5\baselineskip}
\flushleft{$^c$ We only include these filters in our strong lensing model. We approximate $g,r$ and $i$ filters with their SDSS counterparts}\vspace{-0.5\baselineskip}
\flushleft{$^d$ We approximate $VIS$ magnitudes as $VIS = (r+i+z)/3$ (S. Niemi, Private communication)}
\flushleft{$^e$ There is some variation in sky-brightness due to zodiacal and scattered light, but we adopt 22.2 as the $VIS$ sky brightness for all Euclid pointings. { We also neglect the fact that half of the Euclid area will have 2150s total exposure time due to the dithering pattern. (S. Niemi, Private communication)}}
\end{table*}

\section{Defining a detectable lens}
\label{sec:detectability}
Having generated a population of strong lenses and simulated observations of them, we have $\sim$10$^9$ exposures. Whilst each exposure contains a simulated strong lens, not all of them can plausibly be detectable as a strong lenses. Systems with very low signal-to-noise or where the Einstein radius is much less than the seeing are unlikely to be distinguishable as strong lenses. Robotic strong lens finders already exist \citepeg{haggles}, but require a large amount of survey specific tuning and a realistic training set. Most imaging-based strong lens searches require at least some level of human input. Human classifications introduce a stochastic element \citepeg{ringfinder}, but a human expert tends to look for defining features when assessing the probability of strong lensing. The patterns a human lens finder looks for are typically arc-like features, the presence of a similar-color counter image on the opposite side of the lens, features that are a different color (typically bluer) to the putative lens galaxy, and a morphology that is plausible fit by a strong lens model. Systems that meet all of these criteria are typically classified as A-grade strong lens candidates; B-grade candidates often show arc-like features without an obvious counter image, or multiple same-color sources too small to show tangential shearing. The final confirmation of strong lensing typically requires either spectroscopic confirmation of the source redshift, or a compelling lens model. Whilst spectroscopy is expensive, we optimistically assume that any system obviously featuring detectable arcs is a discoverable strong lens, given that all of our systems can be fitted with a strong lens model. 

For an SIS strong lens, the center of the counter images are separated by twice the Einstein radius the magnification is purely tangential \citep{SEF}. Thus the quadrature sum of the seeing and twice the source size must be less than twice the Einstein radius, or counter images will be unresolved. For tangential shearing to be observable the source must be resolved in the tangential direction, i.e. the product of the source size and the magnification must be greater than the seeing, additionally the magnification must be sufficiently large that the source is noticeably sheared. Based on visual inspection of lensed galaxy subtracted residuals we assume that a magnification of at least three and a signal to noise of at least 20 is required to unambiguously recognize arcs. 
 By definition our strong lenses show a morphology consistent with strong lensing, but we do not insist that the source and lens be significantly different colors as color information may not be required for future lens finders \citepeg{joseph}.

Our criteria for a detectable strong lens are that the center of the source is multiply imaged
\be
\label{eq:lenscriteria1}
\theta_E^2>x_s^2+y_s^2,
\ee
where $\theta_E$ is the Einstein radius and $(x_s, y_s)$ are the unlensed source position. The image and counter-image must be resolved
\be
\label{eq:lenscriteria2}
\theta_E^2>r_s^2+s^2/2,
\ee
where $s$ is the seeing and $r_s, (x_s, y_s)$ is the unlensed source size. 
Tangential shearing of the arcs must be detectable;
\be
\label{eq:lenscriteria3}
\mu_{\mathrm{TOT}} r_s > s, \;   \mu_{\mathrm{TOT}} > 3.
\ee
where $\mu_{\mathrm{TOT}}$ is the total magnification of the source. We also require that the source be detected with sufficient signal to noise for a human to recognize that Equations \ref{eq:lenscriteria1}, \ref{eq:lenscriteria2} and \ref{eq:lenscriteria3} are satisfied. 
\be
\label{eq:lenscriteria4}
\mathrm{SN}_{\mathrm{TOT}}>20.
\ee
where $\mathrm{SN}_{\mathrm{TOT}}$ is the total signal to noise of the lensed residual.

We insist that all detectable lenses obey Equation \ref{eq:lenscriteria1}  and that Equations \ref{eq:lenscriteria2} through \ref{eq:lenscriteria4} be satisfied in at least one imaging band. This is more constrictive than the simplest definition of a strong lens (multiple imaging of some part of the source), however we would be surprised if future lens finders can discover a significant population of galaxy-galaxy lenses that do not pass our selection cuts. The one class of galaxy-galaxy lenses that will be discoverable but does not pass our criteria are systems with an unresolved source whose image morphology is still sufficient to detect lensing i.e. quadruple image systems with sufficiently large Einstein radii and a sufficiently blue source to resolve the images from the lens galaxy; this population is likely to be small.

\section{CFHTLS: testing the model against a well studied lens sample}
\label{sec:CFHT}

\begin{figure}
  \centering
    \includegraphics[width=\columnwidth,clip=True]{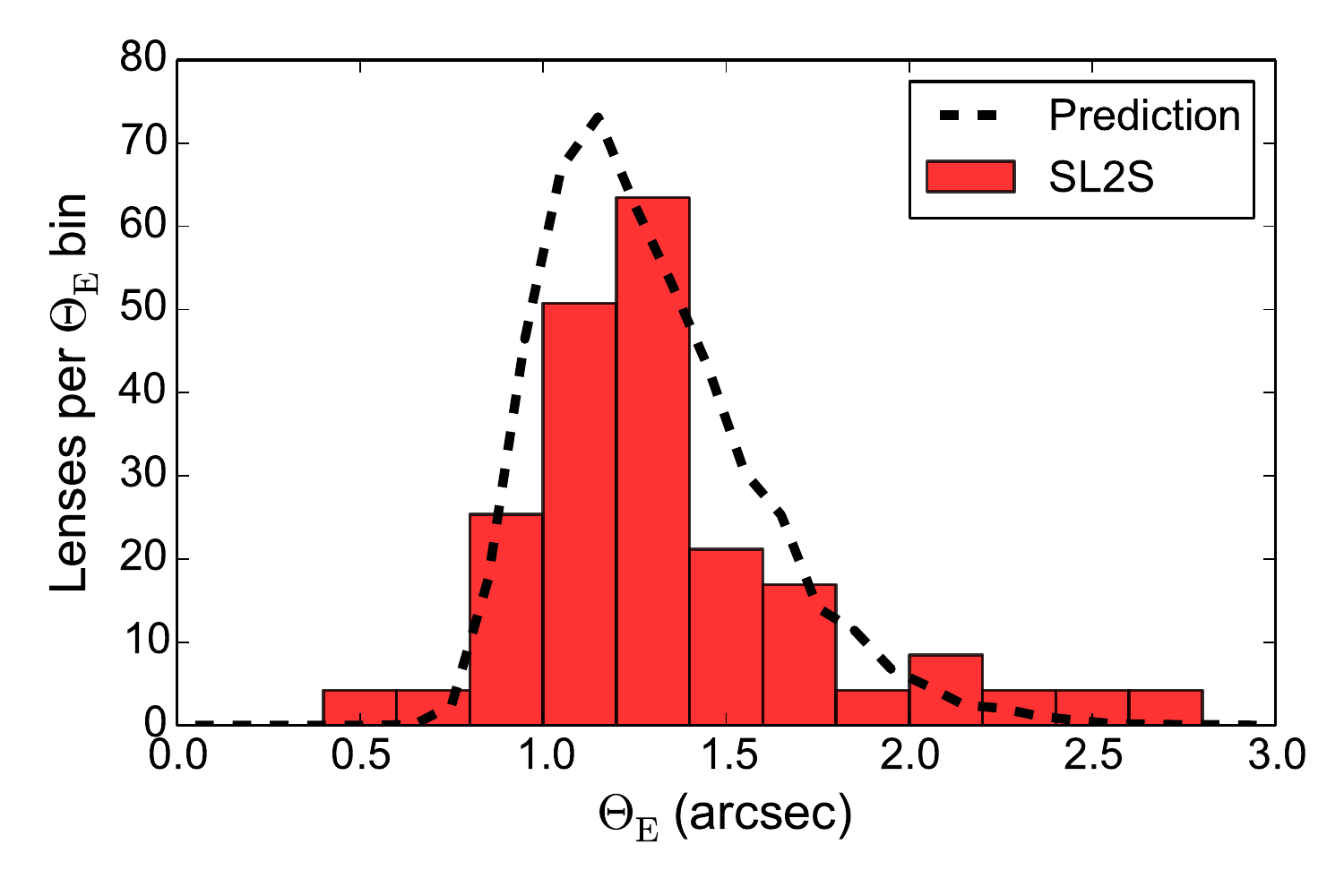}
    \caption {The Einstein radius of the SL2S galaxy-scale lens sample, as fit by \citet{sonnenfeld3} shown in the gray bars and the Einstein radii distribution forecast by our model shown by the black dashed line, assuming a \ringfinder--like search of our simulated SL2S survey. The area under the black dashed line corresponds to 230 lenses. The gray-bars have been scaled to have a total weight of 220 systems, equally up-weighting each of the 52 lenses modeled by \citet{sonnenfeld3} to match the predictions of \citet{ringfinder}.}
    \label{fig:reinpred}
\end{figure}

The CFHTLS has been extensively searched for strong lenses, both visually \citep{spacewarps2} and semi-automatically \citep{ringfinder,more} yielding approximately a hundred confirmed lenses ranging from galaxy to group scales. Our simulations generate the types of lenses that form the SL2S galaxy-scale sample of \citet{sonnenfeld3}, discovered by \citet{ringfinder} using the \ringfinder algorithm. Applying \ringfinder to the real data yielded 401 candidates of which 49 are spectroscopically confirmed. Based on incomplete follow-up \citet{ringfinder} predicts 220 discoverable lenses in CFHT. Applying our detectability criteria to our simulated CFHTLS co-adds, we forecast 800 discoverable strong lenses. However, after applying the PSF cross-convolution and $g - \alpha i$ difference imaging approach of \citet{ringfinder} we would only classify 290 of our candidates as still discoverable.  Roughly half of lost systems are due to the poorer seeing in the $g - \alpha i$ difference image and the other half are lost due to the degraded signal-to-noise. After applying the lens galaxy cuts of \citet{ringfinder}, the number of detectable lenses decreases to 230 systems, in good agreement with the results of \citet{ringfinder}. 
We should not expect our forecast and  the  CFHT results to agree this well; \ringfinder can find lenses with $\mu<3$ if the source is intrinsically tangential to the lens, and does not have strict criteria on total signal-to-noise or on the seeing, both potentially increasing the discoverable lens population. Conversely, some of our simulated lenses that a human would consider obvious candidates are missed by the automated part of \ringfinder. This was  found by \citet{ringfinder} where 77 serendipitously discovered probable lenses during the development of \ringfinder that were not part of the final statistical sample of 330 candidates. 

In Figure \ref{fig:reinpred} we compare the Einstein radii of the forecast 200 lenses with the values inferred by \citet{sonnenfeld3} from modelling the SL2S sample and show that the distribution of Einstein radii are similar, thus our seeing criteria (Equations \ref{eq:lenscriteria2} and \ref{eq:lenscriteria3}) is reasonably representative of what is achievable with an automatic ring-finder.

Given the similarity of both the number of discoverable lenses and the Einstein radius distribution with that found in \citet{sonnenfeld3} and \citet{ringfinder} we are confident that our lens population and detectability criteria are good representations of what is discoverable in the real Universe. In the rest of this work we will report the number of lenses detectable in one band assuming Poisson limited galaxy subtraction (800 in the CFHTLS co-adds) as a target for future lens finders. We will also report the number of lenses discoverable in the $g - \alpha i$ difference image, although we will not apply the \ringfinder cuts on lens redshift and $i$-band magnitude and assume the PSF of the difference image can be set to the worst of the $g$- or $i$-band PSFs rather than cross-convolving the PSFs. These seem plausible upgrades to \ringfinder and gives a forecasts of 480 lenses discoverable in the CFHT co-adds. This forecast represents a more pessimistic (but more achievable) view of where lens finding is likely to be in the next decade.

\section{DES, LSST and Euclid: On-going and future lens surveys}
\label{sec:future}

\begin{table*}
 \caption{The predicted number of discoverable lenses using the model described in this work applied to current and future large area surveys.
\label{table:results}} 
 \begin{tabular}{@{}lcccccc}
  \hline
Survey       &      \multicolumn{2}{c}{Full Stack}    &\multicolumn{2}{c}{Best Epoch}    &\multicolumn{2}{c}{Optimal Stack}\\                
Criteria$^a$ &  $g$,$r$ or $i$& $g-\alpha i$&  $g$,$r$ or $i$& $g-\alpha i$&  $g$,$r$ or $i$& $g-\alpha i$ \\                
 \hline
DES    & 1400 & 800 & 800  & 340 & 2400 & 1300 \\     
LSST$^b$& 39000  & 16000  &  17000 & 3000 & 120000 & 62000\\     
Euclid$^c$&170000  & 0$^d$ &  \\      
 \hline
 \end{tabular}
\newline\footnotesize
\flushleft{$^a$ The detectability criteria applied; for the first column is that discoverable systems must satisfy Equations \ref{eq:lenscriteria1} through \ref{eq:lenscriteria4} in at least one band assuming Poisson noise-limited galaxy subtraction, whilst for the second column the criteria must be satisfied in the $g-\alpha i$ difference image with the PSF of the $g$ and $i$ images matched to have the same seeing. The true number of lenses that forthcoming surveys will discover is likely to be between these two values.}
\vspace{-0.5\baselineskip}
\flushleft{$^b$ Our source population is insufficiently deep to include high magnification lenses with intrinsically ultra-faint ($i>27$) sources. The numbers in the full and optimal stack columns may therefore be underestimates -- see Section \ref{sec:sourcepoperrors}.}
\vspace{-0.5\baselineskip}
\flushleft{$^c$ Since we assume all Euclid exposures have the same PSF and background, stacking all the exposures gives maximal signal to noise.} 
\vspace{-0.5\baselineskip}
\flushleft{$^d$} Since Euclid has only one visible-light filter, constructing an optical blue minus red image will not be possible.
\end{table*}

We now consider the number of detectable lenses in the DES, LSST and Euclid wide surveys. In the final co-added images of each survey we forecast that DES can discover 1400 lenses, LSST can discover 39000 lenses and Euclid can discover 170000 lenses. Requiring that the lenses be discoverable in the $g- \alpha i$ image decreases these numbers to 800 and  16000  for DES and LSST respectively. {Since Euclid has only one visible filter it is not possible to construct an optical blue minus red image so no lenses can be found this way; it may be possible to use the near infra-red channels although they have decreased depth and resolution, or to use lower resolution multi-band optical data to add color information to the VIS imaging \citepeg{pansharpening}, but we have not investigated these possibilities.}

\subsection{Alternative strategies for combining multiple exposures}

\begin{figure*}
  \centering
    \includegraphics[width=\columnwidth,clip=True]{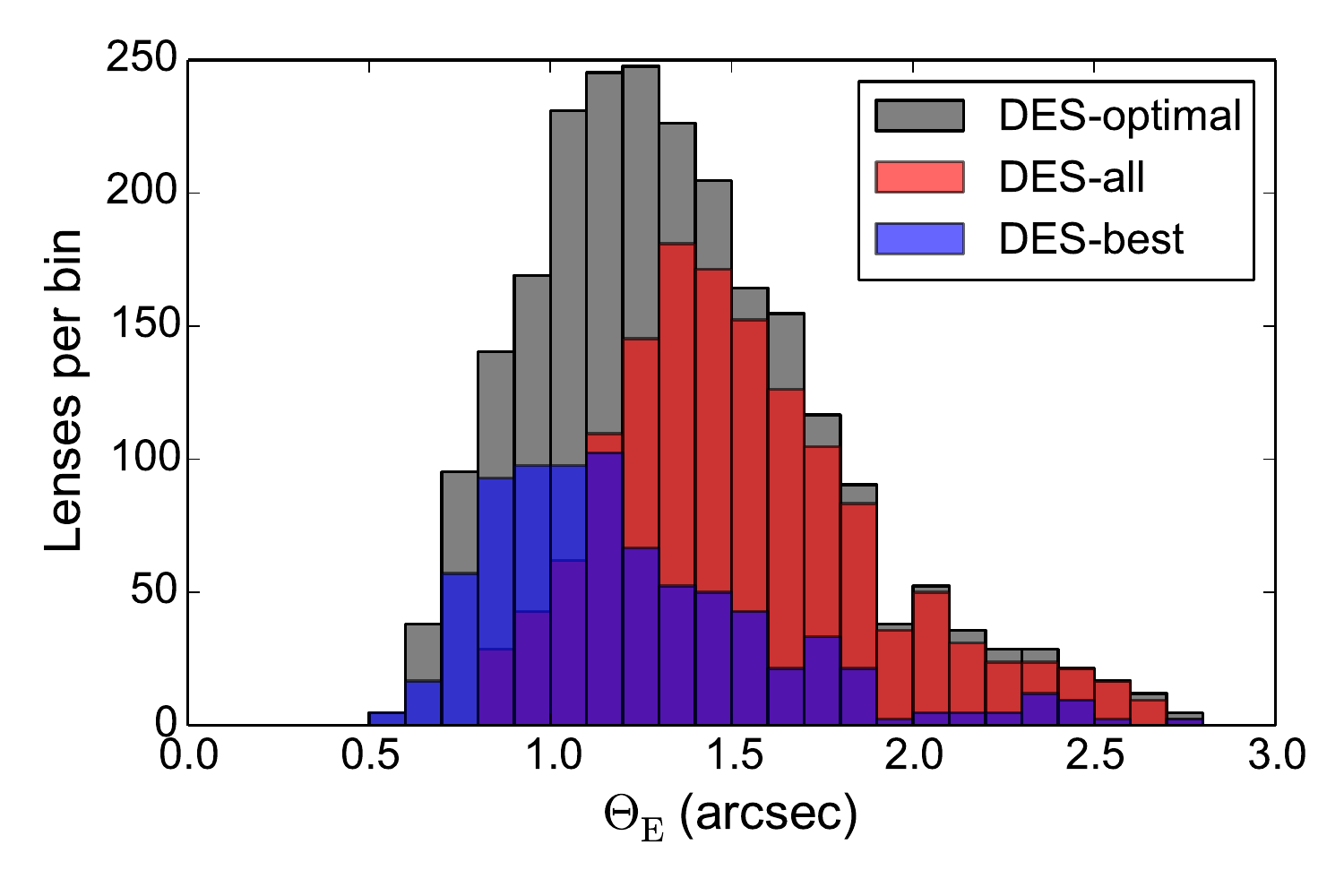}
    \includegraphics[width=\columnwidth,clip=True]{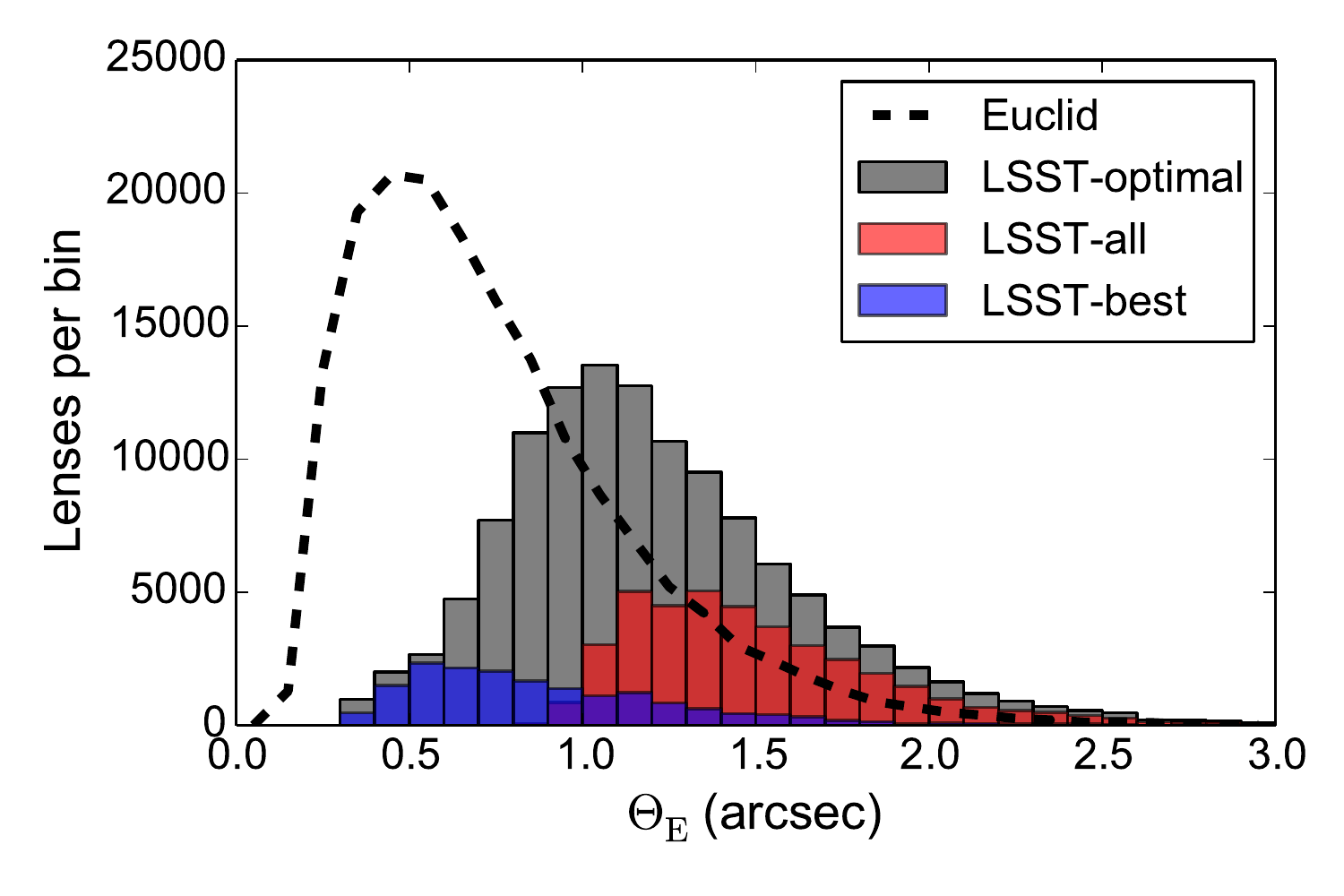}
\caption {The Einstein Radius distribution for lenses discoverable in DES (left panel) and LSST (right panel). The red distributions show the Einstein radius distribution of the lenses discoverable in the full stack of the final survey, whilst the blue distributions show the population discoverable in the best single-epoch imaging. The gray distribution shows the Einstein radius distribution of the lenses that can be discovered if one searches for lenses in all possible (seeing-ordered) stacks for each survey. The dashed black line in the right panel shows the Einstein radius distribution for galaxy-galaxy lenses discoverable in Euclid.}
\label{fig:b}

\end{figure*}

Each object in these surveys will be imaged many times; individual exposures can be combined to make a deeper, stacked image, however for ground based images the exposures will not have comparable seeing; with some exposures having much higher resolution than others. Since our detection criterion requires the seeing to be less than the Einstein radius, adding poor seeing images to the stack can decrease the discoverable lens population. Therefore for DES and LSST we investigate alternative stacking strategies. A trivial alternative strategy to co-addition is to discard all exposures except the image with the best seeing in each filter, in effect trading signal-to-noise for resolution. This strategy yields 800 (17000) detectable lenses for DES (LSST) of which 340 (3000) are still discoverable in the difference image. Fewer lenses are discoverable in the best seeing images than the co-adds, but they are not the same lenses, as Figure \ref{fig:b} shows; the co-add imaging finds the higher mass lenses whilst the best single epoch imaging discovers the systems with the brightest lensed sources.

Given our detection criteria, an optimal stacking exists which only includes exposures where the seeing is sufficiently good for $\theta_E^2>r_s^2+(s/2)^2$ and $\mu r_s > s$ to be satisfied. Since the source size and Einstein radius is not known a-priori performing this optimal stacking on a real data set is impossible, but is still physically meaningful: applying a lens finder to all possible stacks (ordered by seeing), only increases the search effort by a factor of the number of exposures per filter and naturally includes the optimal stack. Alternatively a lens finder that can simultaneously model all the individual exposures without any stacking, could potentially outperform even this stacking strategy. With optimal stacking, and perfect galaxy subtraction DES (LSST) can discover 2400 (120000) lenses; 1300 (60000) are still detectable in the $g- \alpha i$ difference imaging. For LSST, 70000 -- more than half -- of the 120000 lenses that are discoverable in the filtered stack will not be discoverable in either the best single epoch imaging or the final co-add.

\subsection{The properties of the lens populations}

\begin{figure}
  \centering
    \includegraphics[width=\columnwidth,clip=True]{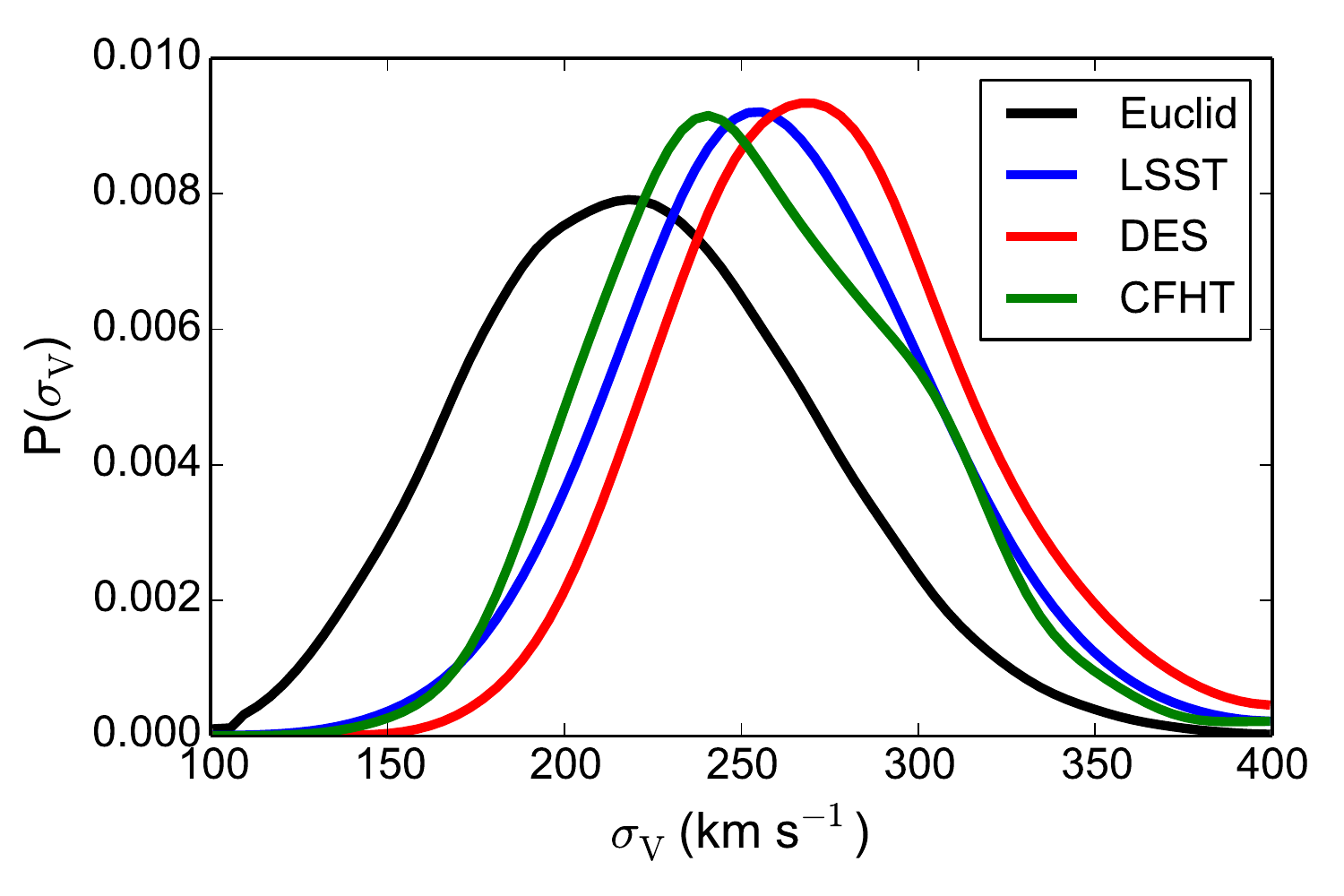}
    \caption {The velocity dispersion distribution for the lenses discoverable in forthcoming wide area surveys. Black shows Euclid, green shows CFHT, blue shows LSST and red shows DES. Surveys with better seeing are able to discover lenses with lower Einstein Radii and hence probe intrinsically less massive objects.}
    \label{fig:sig}
\end{figure}

\begin{figure}
  \centering
    \includegraphics[width=\columnwidth,clip=True]{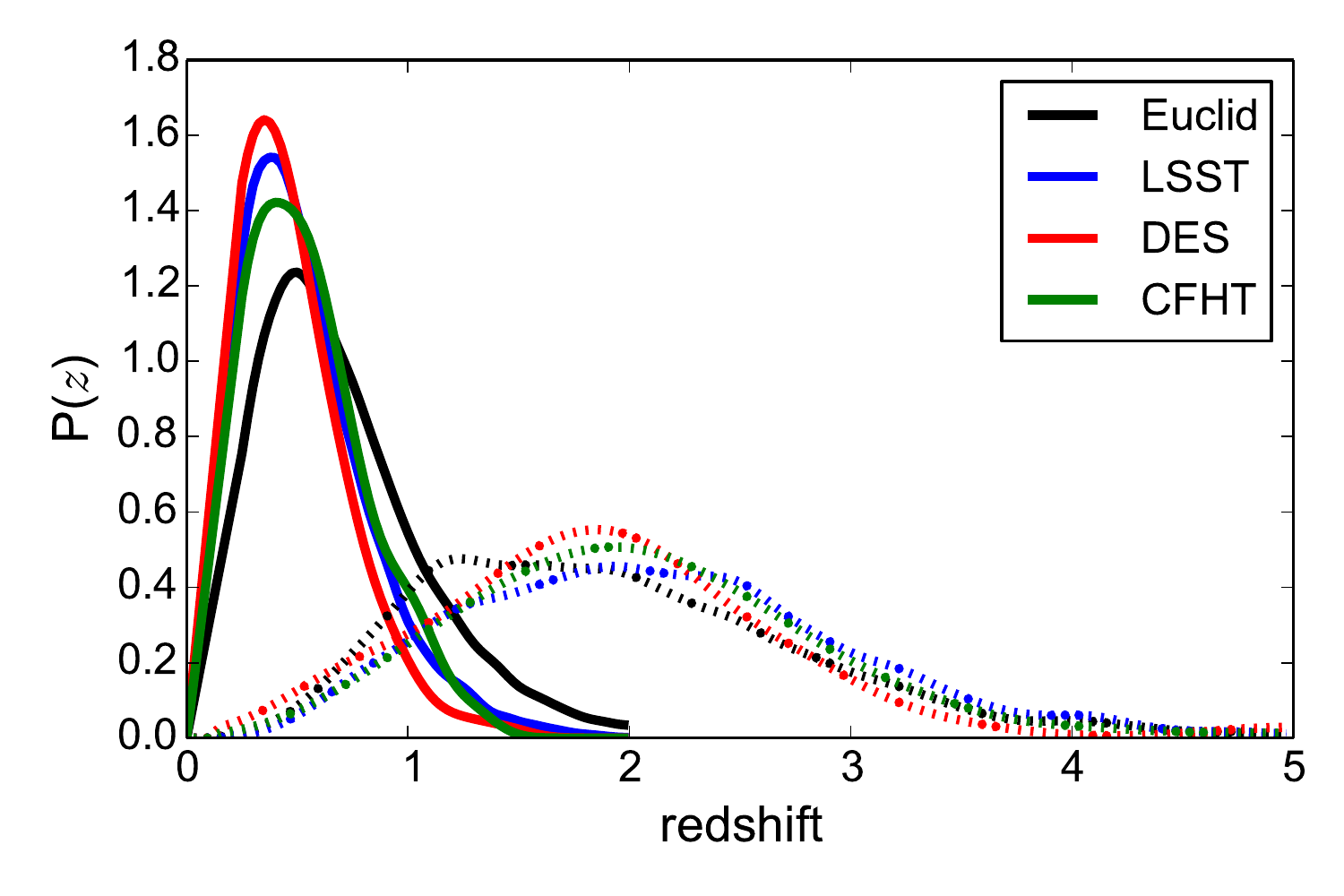}
    \caption {The redshift dispersion distribution for the lenses discoverable in forthcoming wide area surveys. The lens redshifts are shown by the solid lines, the source redshifts are shown by the dashed lines. Black shows Euclid, green shows CFHT, blue shows LSST and red shows DES.}
    \label{fig:zdist}
\end{figure}

Our model allows us to forecast more than the total number of discoverable strong lenses, it also allows us to forecast the properties of the population. This allows us to predict if forthcoming surveys will open up new regimes of strong-lens science. A-priori we expect that the high resolution imaging of Euclid opens up the possibility of discovering a large number of low-mass strong lenses and the large area and depth, but comparatively poorer seeing, of DES and LSST will make them efficient at finding massive lenses. We see both these expected effects in figure \ref{fig:sig}, where we plot the predicted velocity dispersion function distribution of each survey.

Figure \ref{fig:zdist} shows the source and lens the redshift distributions of discoverable lenses in each of the surveys. There is no significant variation in the distribution of source redshifts between the surveys, although this is possibly due to the fact that our model only includes the $g,r$ and $i$-band filters for CFHT, DES and LSST which precludes large numbers of high redshift sources and Euclid's shallowness will also precludes finding a high redshift source population without extreme magnification. An analysis of the discoverable infra-red lens population, may see a more significant variation of source redshift with imaging depth. Despite being the shallowest survey considered, we find that Euclid finds the highest fraction of high redshift lenses (Figure \ref{fig:zdist}). For an isothermal lens, the Einstein radius is proportional to $D_{ls}/D_s$ hence at fixed source redshift, higher redshift lenses have a smaller Einstein radii making them harder to resolve in ground-based imaging. However, since the highest redshift lenses currently known are at $z \sim$1.6 \citep{wonghighz,vanderwel}, {Figure \ref{fig:zdist} suggests that both DES and LSST are also capable of discovering higher redshift lenses than currently known.}

\section{Discussion: Errors on Forecast numbers}
\label{sec:errors}

Our forecasts are based on a well-motivated prescription for generating a deflector and source population, lensing the background source, simulated observations of the lens and criteria that define a detectable lens. All of these stages have required simplifying assumptions that have the potential to systematically bias the lens number. In this section we investigate how small perturbations to the model can affect the forecast number of resolved lenses discoverable in the optimally filtered $g-\alpha i$ filtered stacks of the mock DES survey. 

\subsection{Deflector Population}

We have assumed that there is no evolution in the co-moving number density of deflectors and that the velocity dispersion function does not evolve with redshift. This is consistent with the results of \citet{bezanson} and \citet{shu}, which show minimal evolution of the VDF out to z=0.5. Measuring velocity dispersions at higher redshifts is observationally challenging, but it is possible that there is evolution of the massive galaxy population at higher redshifts. 
\citet{mason} investigated the evolution of the VDF using the stellar mass as a tracer, fitting a model of the form $\log(\sigma/\sigma_{z=0}) \propto (1+z)^\beta$ and co-moving number density evolving as $\propto (1+z)^\alpha$, and found $\alpha=2.46 \pm 0.53$ and $\beta = 0.20 \pm 0.07$. For $\alpha=2.46,\beta = 0.2$, the number of lenses detectable in the DES $g-\alpha i$ difference images increases by sixty percent. However, the magnitude of this uncertainty is most likely overestimated, as the fits of \citet{mason} are skewed by evolution at high redshift where our model already predicts no significant lens population. Given the evidence for no significant evolution out to $z\sim$1.5 \citep{bezanson, shu} the no-evolution model seems a much more realistic representation of the Universe.
We also investigated how deflector ellipticity impacts the number of detectable lenses, and saw no significant change in the discoverable lens fraction when increasing the ellipticity of all lenses by ten percent. 

\subsection{External Shear}

Strong lenses do not tend to live in isolation; elliptical galaxies tend to cluster hence the tidal field of nearby masses perturbs strong lens systems. Simulating the full environments of strong lenses is beyond the scope of this work, but we can investigate how significant this effect is using a reasonable toy model. Our toy model assumes that the two components of the shear $\gamma_1, \gamma_2$ are independently drawn from a Gaussian of width $\sigma=0.05$; this effect is larger than would be expected for random lines of sight \citepeg{pace} but is in reasonable agreement with the external shears found by modeling strong lenses \citepeg{wong}. Applying this toy model we find a five percent increase in the number of detectable galaxy-galaxy strong lenses in the DES $g-\alpha i$ difference images; this moderate shear effect on the selection function implies that the comparatively large shears in photometrically selected galaxy-galaxy strong lenses is typical of the environment of elliptical galaxies rather than a selection bias.

\subsection{Source population}
\label{sec:sourcepoperrors}
\begin{figure}
  \centering
    \includegraphics[width=\columnwidth,clip=True]{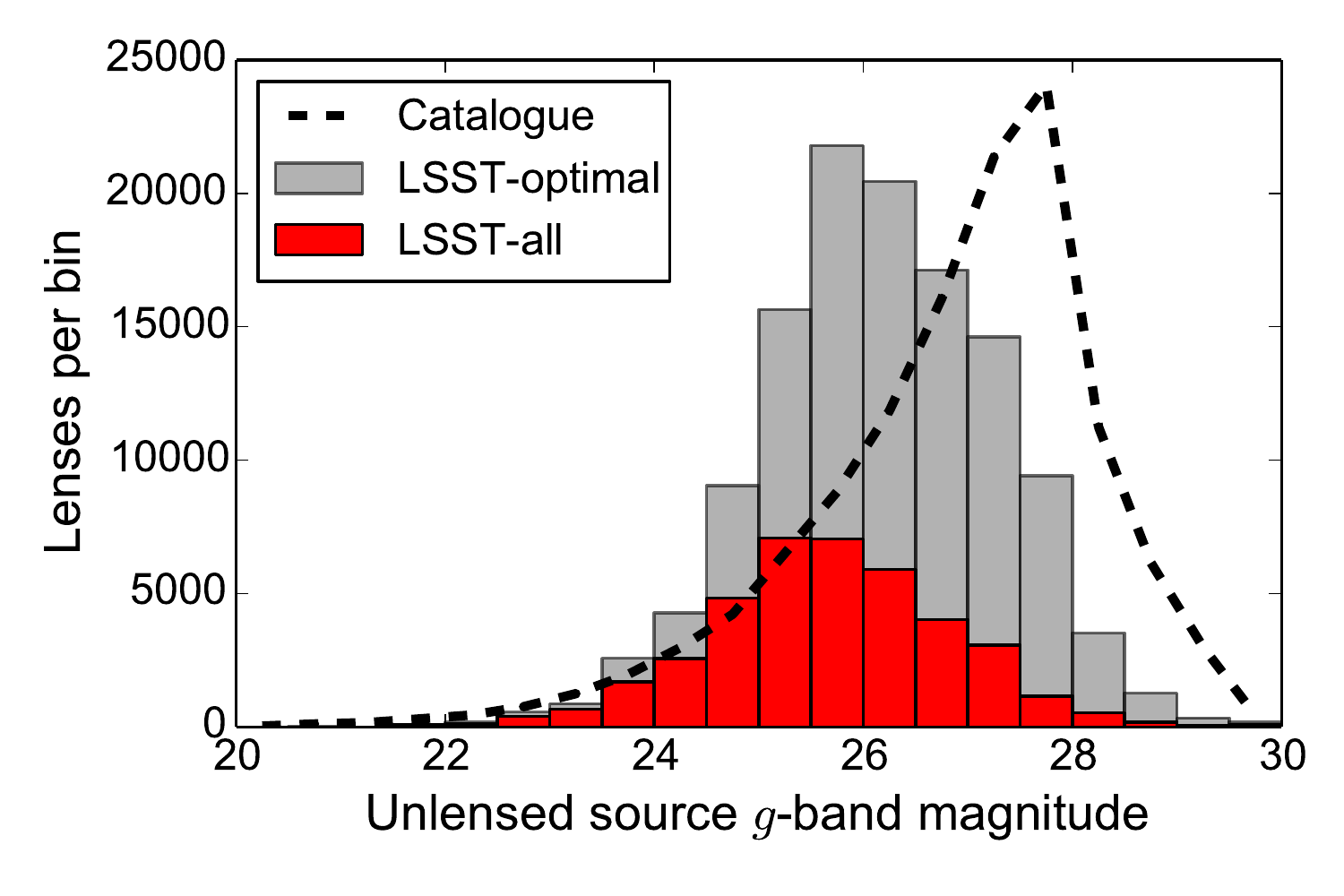}
    \caption {The distribution of unlensed source g-band magnitudes for strong lens systems discoverable with LSST. The red histogram shows the population discoverable in the full LSST stack, whilst the gray population are discoverable if one searches for lenses in all possible (seeing-ordered) stacks. The shape of the distribution of $g$-band magnitudes for the source catalog \citep{connolly} is shown as the dashed black line. Most lenses discoverable with LSST are intrinsically much brighter than the completeness limit of the source catalog. There is therefore unlikely to be a large population of discoverable strong lenses whose sources are too faint to be included in the \citet{connolly} catalog.}
    \label{fig:lsstg}
\end{figure}

 Our LSST forecasts are underestimated, since our unlensed source catalog is of comparable depth to the LSST co-add. In figure \ref{fig:lsstg} we show that this is not a major effect; for most discoverable lenses the sources are brighter than the completeness limit of the source catalog. Given the luminosity function, this result appears counter-intuitive however faint sources are also small, hence few intrinsically very faint galaxies can be resolved in the LSST imaging. The source catalog has been matched to the number counts of deep surveys, so the number of potential background sources should be robust however there is scope for small changes to the number of detectable lenses, since the source population colors do not  match the observed population \citep{delucia} and the redshift distribution of sources fainter than $i > \sim$24 requires extrapolation.

The COSMOS photometric redshift catalog \citep{ilbert} is complete down to $i=25$ and has multi-band HST photometry; this catalog implicitly includes the complex distribution of source colors and hence provides a more realistic representation of the bright source population. Using the COSMOS photometric redshift catalog, we predict 760 lenses discoverable in the DES $g-\alpha i$ difference images. The original model had 840 lenses with unlensed source magnitudes of $i<25$. Redshift catalogs much deeper than $i=25$ do not exist, hence the redshift distribution of faint sources is poorly constrained; the uncertainty on the number of lenses with intrinsically faint sources is likely to be larger than the ten percent found for sources with $i<25$, but we cannot precisely quantify this.

The intrinsic sizes of the unlensed source population is a key parameter in predicting the detectable lens population, since it sets the maximum magnification and the seeing required to resolve that the source is arced. Most of our sources are drawn from a high-redshift faint population that is poorly understood. \citet{shibuya} recently analyzed the median sizes of galaxies in HST from redshift 0-10, finding that it is proportional to $(1+z)^{-1.12}$; if we use this result rather than the $r_s \propto (1+z)^{-1.2}$ assumed in Equation \ref{eq:size}, we find that this makes sources somewhat larger at high redshift (9 per cent at $z=2$), which decreases the DES detectable lens population by six percent. An additional effect is related to the morphology of the sources, which are likely highly irregular. Our assumption of an elliptically symmetric exponential profile for the source is overly simplistic and if most of their flux originates from high density knots of star formation, their peak surface brightness may be larger and hence they may be more easily detected than our simple model predicts.

\subsection{Definition of Detectability} 

Whilst our detectability criteria are well matched to the CFHT lens sample, it is possible that lens finding in future surveys will be more difficult; pushing lens finding to fainter sources and smaller Einstein radii may introduce more false positives, forcing future lens finders to be more selective about what they consider to be a good lens candidate. We investigate three possible tightenings of the detection criteria; changing the total signal-to-noise threshold from 20 to 40, changing the magnification threshold from 3 to 4 (i.e. requiring a more curved source) and changing the resolution threshold from $\theta_E^2>r_s^2+(s/2)^2$ to $\theta_E^2>r_s^2+s^2$ (i.e. requiring arcs and counter-images to be better resolved from each other and the lens). The number of lenses detectable in the DES $g-\alpha i$ difference images halves (down to 600) for the change in signal-to-noise, decreases by fifteen percent for the increased magnification constraint, and decreases by forty percent when requiring the lensed features to be more easily resolved. The change in signal-to-noise has the most significant effect, but for visual inspection of residuals, lensed features that are tangentially resolved and detected at a signal-to-noise of 20 are already unambiguously strong lensing.

%
%
\section{Conclusion}
\label{sec:conclude}

In this paper we have investigated the question {\it "how many galaxy-galaxy strong lenses can forthcoming surveys discover?"} We have developed a realistic model for the lens and source populations to build a population of lenses that is representative of the strong lenses in the real Universe. We have then simulated the observations of forthcoming surveys, paying particular attention to the seeing distributions in each observing band - which we find to be a key variable in the number of detectable strong lenses. We have then analyzed these simulated observations to investigate which of the systems are plausibly detectable, finding that in our most optimistic scenario DES, LSST and Euclid can discover 2300, 120000, and 280000 lenses respectively, but only if the lens galaxy can be subtracted without degrading the resolution and signal-to-noise of the sample. We have shown that the detectability of lenses depends strongly on the search strategy adopted; despite the fact that lensed sources are typically much bluer than the lens, $g- \alpha i$ difference imaging such as implemented in \citet{ringfinder} only finds around a quarter of the optimistically discoverable lenses since it degrades both the PSF and the signal-to-noise. If future lens searches still use $g- \alpha i$ difference imaging but can match the PSFs to the worst of $g$ or $i$ seeing this can potentially discover twice as many lenses as the current foreground subtraction method implemented in \ringfinder; such a finder could potentially discover 1300 lenses in DES and 62000 in LSST. The absence of color information will require the development of single band finders \citepeg{joseph} to find lenses in Euclid.

For LSST (and a lesser extent DES) we found that the majority of discoverable lenses can neither be discovered in the full co-add nor the best single epoch-imaging; the interplay between signal-to-noise and seeing mean that all possible (seeing-ordered) stacks must be searched to maximize the number of discoverable lenses. 

We have shown in Section \ref{sec:errors} that the dominant source of error is the definition of a detectable lens. In future, our definition of detectability can be re-calibrated by running lens finders on our simulated lenses like; this will allow us to robustly understand the selection effects of different lens-finders. Section \ref{sec:errors} shows that the model itself appears to be accurate to around a few parts in ten; given the uncertainty on defining detectability, this is sufficient for the purpose of this work, but going beyond this level of accuracy will need a significant improvement in our understanding of the faint source population and redshift distribution.

One major issue that we have not addressed here is the question of false positives. The detectaility criteria outlined in Section \ref{sec:detectability} seem reasonable for current datasets but may prove optimistic in future if greater depth and improved seeing reveals a large number of non-lenses that have morphologies similar to lenses. {In particular, we have not required the detection of counter images, which may be necessary to achieve an acceptable sample purity. It may also be that some regions of parameter space produce less pure lens samples than others; for example it is easier to recognize tangential shearing in thinner arcs; lensed low surface brightness sources may be hard to distinguish from face-on spiral arms or tidal features of a candidate lens galaxy. Spectroscopic confirmation of lens candidates is expensive and robust lens modeling presents a challenge in the face of large numbers \citep{brault}; future searches may require much higher purity than current samples. The selection function of future lens finders may be more conservative than we have assumed in this work.}

We have made the code used in this work publicly available\footnote{Code is available at \url{github.com/tcollett/LensPop}}; future strong lens finding algorithms can therefore be tested on these systems, to asses what subset of the full lens population they will discover. If --once the finders has been run on the simulated lens sample-- the properties of discovered lens populations deviate significantly from that predicted by our model it will provide evidence that some aspect of the model's deflector or source population is incorrect. Whilst strong lensing has already shed light on the high redshift source population \citep{slacsXI}, and the mass distribution in strong lens galaxies \citepeg{auger2010, sonnenfeld4}, these results may be biased by the discoverable strong-lens selection function. Given enough lenses and computational power a model such as the one presented in this work could plausibly be used to infer the many parameters of Section \ref{sec:population}. This would implicitly include the strong lens selection function \citepeg{sonnenfeld5} and should give unbiased inference on the properties of the deflector and source populations.

Over the next decade, surveys will have sufficient signal-to-noise and resolution to discover hundreds of thousands of galaxy-galaxy strong lenses, however no one human can hope to perform visual inspection on such a large sample. Since all current lens finders need some level of human inspection and extracting scientific results from lenses requires human input in the form of lens modeling, Making full use of these hundreds of thousands of lenses will require a huge improvement in automation \citep[e.g.]{haggles, brault} or crowd-sourcing \citep{spacewarps1,spacewarps2,kueng}.


\section*{Acknowledgements}
The simulated images and code used to generate them are freely available upon request from the author and from \url{github.com/tcollett/LensPop}.

I am grateful to David Bacon, Bob Nichol, Phil Marshall and Anupreeta More for useful discussions and comments on the draft manuscript. I am greateful to the anonymous referee for helpful comments. I am grateful to Matt Auger for the lensing code and the color image code used to generate Figure \ref{fig:examplelens}.
I also thank Chris D'Andrea, Sami Niemi and Lynne Jones for providing information on the observing capabilities of DES, Euclid and LSST respectively. This work has used the LSST operations and sky simulations; I am grateful to  Andy Connolly for sharing these and to Scott Daniel and Gary Burton for enabling access to them.

\comments{
\appendix
}




\label{lastpage}

\end{document}